# A Concrete Example of Inclusive Design:Deaf Oriented Accessibility

Claudia SavinaBianchini[1], Fabrizio Borgia[2], Maria De Marsico[2]

[1]Laboratoire EA3816-FoReLL, Poitiers, France

[2]Sapienza University of Rome, Rome, Italy

## Abstract

One of the continuing challenges of Human Computer Interaction research is the full inclusion of people with special needs into the digital world. In particular, this crucial category includes people that experience some kind of limitation in exploiting traditional information communication channels. One immediately thinks about blind people, and several lines of research aim at addressing their needs. On the contrary, limitations suffered by deaf people are oftenunderestimated. This is often the result of a kind of ignorance or misunderstanding of the real nature of their communication difficulties. This chapter aims at both increasing the awareness of deaf problems in the digital world, and at proposing the project of a comprehensive solution for their better inclusion. As for the former goal, we will provide a birds-eye presentation of history and evolution of understanding of deafness issues, and of strategies to address them. As for the latter, we will present the design, implementation and evaluation of the first nucleus of a comprehensive digital framework to facilitate the access of deaf people into the digital world.

# A concrete example of inclusive design:Deaf Oriented Accessibility

## 1. Introduction

Tremendous technological advances have allowed to dramatically improve Human Computer Interaction (HCI). In particular, the expressiveness of interfaces to access and use digital services and resources has increased, while their difficulty for the final user has decreased over time. Communicating with a computer has evolved from using punched cards and command line interfaces to the so defined “natural interfaces”. This can be compared to the evolution from free climbing by ropes to the elevator and perhaps to the rocket to reach high targets. In fact, designing User Interfaces (UIs) developed into designing a more comprehensive User Experience (UX). Notwithstanding this, and the apparent global access to any kind of digital equipment and resource, research and practice in HCI still face many challenges. One of its most valuable achievements from the social point of view would be the full inclusion in the digital world of people with special needs. At present, we can still assess the presence of two main gaps giving raise to digital divide: a technological one and a “sociological” one. The former refers to the difficulty of accessing some enabling technology, e.g., wide bandwidth channels, which can still be observed in unsuspected settings. The latter refers to the difficulties still experienced by specific categories of users. First of all, it is worth reminding the distinction by Prensky(2001) between “ natives and digital immigrants” that still holds in many social groups and refers to the difficulty of adapting to new technologies by those who were born and grew before their introduction.A further affected category includes people experiencing some kind of limitation in exploiting traditional human communication channels. This not only includes disabled people, but also elder users whose perceptive abilities may have been degraded by age.

Since the late ‘90s, as a result of the generalization of Internet, lists of best practices to facilitate access to the network have been developed and formalized. The main attempts to address general accessibility issues have been made by the World Wide Web Consortium (W3C), with the Web Content Accessibility Guidelines (WCAG) document. This document aims at anticipating the needs of people with disabilities, to enable them to enjoy freely and independently the contents of Internet. In fact, WCAG2.0 states in its own abstract that “following these guidelines will make content accessible to a wider range of people with disabilities, including blindness and low vision, deafness and hearing loss, learning disabilities, cognitive limitations, limited movement, speech disabilities, photosensitivity and combinations of these.” (Caldwell et al., 2008).

Blind people and their needs are the group that mostly appears in research addressing accessibility. However, limitations suffered by the deaf[1] are often underestimated and therefore scarcely tackled. This often depends on a kind of ignorance or misunderstanding of the real nature of their communication difficulties. Even in recent years, many HCI researchers still thought that it was sufficient to substitute audio with text, or to use subtitles in videos, to address usability needs of deaf people. We will point out how this caused a lack of attention with respect to very specific problems, which must be addressed in a very specific way. In many cases changing the channel is not sufficient, but it is also necessary to rethink the structure of the information (see for example the e-learning platform designed expressly for deaf people in Bottoni et al., 2013). The digital world offers unique occasions for integration, but people with special needs risk losing them. This risk is especially concrete when such needs are scarcely appreciated, as it happens for deaf. This chapter aims at both increasing the awareness of deaf problems in the digital world, and at proposing the project of a comprehensive solution for their better inclusion. Therefore, we will start by providinga short history of the evolution of understanding of deafness issues, and of strategies to address them. Afterwards, we will briefly present the design and evaluation of the first nucleus of SWORD (SignWriting Oriented Resources for the Deaf), a comprehensive digital framework whose final goal is tosupport deaf people in exploiting the opportunities of the digital world.

[1]Deaf: despite what some people - not working closely with the deaf - may think, there is nothing offensive in the word “deaf” when used to speak of the people belonging to the linguistic and cultural minority called “Deaf Community”; for this reason, we have decided to use this term throughout the chapter.

**2. Vocal Languages, Sign Languages and Accessibility**

Despite deafness appears in the second place in the list of disability considered by the WCAG and mentioned above, we find a late attention to deaf-specific problems. WCAG1.0 guidelines (1999) deal mostly with labeling and transcription of audio content, leaving out alternatives related to Sign Languages(SLs). The only mention relates to a possible management of videos. In practice, the suggestion is to complement the audio content with subtitles and, if possible, with atranslation into SL. Of course, these guidelines are dictated by a sincere effort to solve the accessibility problems of the deaf. However, theypartly reveal the misunderstandingwhich takes to assume that a written transcription of audio content is enough to support deaf users. As we will underline, text is strictly connected to verbal languages, and it relies on underlying linguistic structures which are different from those developed by deaf. Therefore, it presents several difficulties for them even in everyday communication. WCAG2.0 (2008) testifies a progressively increasing awareness of deaf needs and better addresses such issues. For instance, the success criterion 1.2.6, whose satisfaction is necessary to get the highest level of compliance (AAA), states that “Sign language interpretation is provided for all prerecorded audio content in the form of synchronous media types.” This opens the debate about at what extent videos or related techniques can really solve all accessibility problems encountered by deaf people.

In any case, WCAG guidelines are insufficient to grant the navigating autonomy that theyaimat achieving.This is explicitly stated in the document: “Note that even content that conforms at the highest level (AAA) will not be accessible to individuals with all types, degrees, or combinations of disability […] Authors are encouraged […] to seek relevant advice about current best practice to ensure that Web content is accessible, as far as possible, to this community” (ibid.), and it is precisely in quest for these best practices that this chapter will be developed.

A starting factor to deal with is the heterogeneity of the population with hearing problems, which can be considered from different points of view: medical, socio-cultural, and linguistic. In the medical perspective, two distinct criteria hold, namely degree and age of onset of deafness. Under the socio‑cultural perspective, it is important to consider the closeness to the “deaf community” and self‑identification as “Deaf”or “hearing impaired”. Last but not least, the linguistic point of view considers mastery of a vocal language（VL) and/or of a Sign Language (SL). Among these pillars there are extreme prototypical cases. On one extreme, we find slightly deaf elderly,who communicate in VL, and consider themselves hearing people yet lacking somehearing. At the other extreme, wefind pre‑lingual deaf ‑i.e., those born or turned deaf before learning to speak ‑ who communicate in SL and feel Deaf.In the middle, we have to consider many shades, e.g., deep pre‑lingual deaf who do not know SL, or mild deafwho use SL. One common element among most of the pre‑lingual deaf is, indeed, the problematic relationship with the VL, in particular with its written form. A report drawn up in 1998 in France (Gillot, 1998) highlighted how 80% of the early deaf is considered almost illiterate, despite having duly attended school; this percentage applies to both signer (using SL) and oralist (using exclusively VL) early deaf. In a nutshell (for more details the interested reader can refer to Goldin‑Meadow, 2001), the problem arises from a lack of exposure of the deaf childto the VL since early age. This creates a gap which, regardless of the possible exposure to the SL, determines a patchy acquisition of the VL.Given the strict cognitive/linguistic relationship, such shortages are then reflected in both the production and interpretation of written text. In fact, deaf people tend to reflect their visual organization of the world over the organization of language. Given this peculiarity, they find significant difficulties in both learning and mastering VLs, even in their written form (Perfetti&Sandak, 2000). Moreover, even if deaf people succeed in mastering a VL, dealing with VL content may prove quite a tiring task for them, unless it is performed for a short time. In fact, it can be observed that most of them prefer to communicate using SLs (AntinoroPizzuto et al., 2010a). Therefore, dealing with the issues of deaf-oriented accessibility using only written VL is quite unrealistic (Borgia, et al., 2014), and any VL-based solution (VL captioning and transcription) to overcome the digital divide rarely solves the problem completely. In other words, captioning-based accessibility design may support the needs of people who become deaf after the acquisition of speech and language (post-lingual deafness). However, issues related to pre-lingual deafness are seldom and poorly addressed.The deaf illiteracy becomes even more relevant for our purpose, given the importance of textual information in most applications, and of written communication in the global network. In fact, despite the regular increase of multimedia content, text remains the main mode of content diffusion over Internet: even on sites, like YouTube,that are especially oriented toward video contents, we can observe that titles, descriptions, comments, and almost all the search tools are mainly based on text.

Summarizing the above considerations, we can suggest that the heterogeneity of the deaf population must be reflected in a variety of choices for accessibility: subtitles for the “late” deaf, and alternative solutions for the pre‑lingual or early deaf, whether signers or oralists. To this latter respect, it is important to take into account the preference for SL by the former, and the difficulties encountered by the latter with the VL, despite this being their only language.Unfortunately, at the best of our knowledge, no universally accepted instrument exists to profile a deaf individual on all of the factors that we mention.However, some interesting examples can be found regarding specific application fields, e.g. e-learning(Salomoni*et al.*, 2007)

Though using the singular expression Sign Language, we are rather referring to a class of languages instead than a unique one. Sad to say, as withVLs, even SLs present national and regional differences in the use of signs. However, given the strong perceptual basis of such languages, people signing differently can still understand each other better and more quickly than those speaking different VLs. Indeed, the latter usually do notunderstand each other at all, andtend to use gestures to help communication. In this work we will be focusing more closely on the deaf signers’ case. First, we will show the peculiarity of SL, the problems arising from the lack of a (universally accepted) written form for it, and why signed videos are not enough to allow for an accessibility comparable to that offered by the script. We will then present different systems created for the graphical representation of the SL and, finally, we will propose our solution. It entails the use of two software tools designed to be part of a more comprehensive framework.

**3. A Close View at Sign Languages**

A SL is a language that uses the visual‑gestural channel to convey the full range of meanings that, in VLs,  arerather expressed through the acoustic‑vocal channel. As stated above, it is not the language used by the totality of the deaf. However, the fact that SL exploits the intact visual channel,and related cognitive and communicative structures,makes it more accessible to them.Failure to attain a more widespread diffusion of SL to the pre‑lingual or early deaf is mostly due to historical and cultural factors (a brief history of deaf education is provided by Stokoe, 1960). In fact, since Antiquity, philosophers have questioned the link between thought and speech, often arguing that the absence of speech corresponds to the absence of mind (hence the recent phrase “deaf and dumb”, in which “dumb” means both “not talking” and “stupid”). Starting with the Renaissance, some tutors began to think about strategies for deaf education. Two schools of thought arose: the one formerly most diffused, namely the “oralist” school, concentrates its efforts on teaching the spoken VLat all costs,training a deaf person to communicate through lipreading and speech, and often at the expense of the content; the second school, the “signing” one, favorstaking advantage of the SL for communication for the the full provision of teaching contents, and often neglects teaching of spoken VLs. Of course, the latteris based on the observation that deaf naturally use a kind of visual-gestural communication system, and on the purpose to exploit it also for education. In the second half of the 18th century, the Abbot de l'Épée founded in Paris the first school for the deaf. It developed a teaching method based on the use of the SL to replace the oral communication, and of written French to interface with the rest of the society. This revolution in deaf education, however, was accompanied by severe criticism from those believing that deaf children should learn to speak, and not just to write. After a century and a half of wrangling, the Milan Congress on Deaf Education sanctioned in 1880 the ban of SL from the education of deaf people, trumpeting the slogan “Signs kill speech”. The consequences of those decisions were more or less heavy depending on the country, but the entire deaf education was touched. For instance, in France, home of the Abbot de l'Épée, it was necessary to wait until 1977 to have the ban removed, until 1984 for the establishment of a bilingual school for just a few deaf students, and until 2005 to have a law on the accessibility of handicapped people, sanctioning the right to use SL. In the US, although unaffected by the Congress of Milan and considered a pioneering country in this matter, SL is recognized at single State level as a “foreign language” or as a “language for education” at best. Discussions between pro- and anti-SL are still on the agenda, and are particularly virulent in contexts without laws about SL recognition (like in Italy). In these cases, many lobbies try to convince the legislature that the recourse to SL prevents VL mastery.

These educational discussions pair with the linguistic diatribes. It was necessary to wait until the ’60s of last century, and for the works of the American linguist William Stokoe (1960),to get the first analysis of the linguistic structure of the SL. Those studies finally let it in the Olympus of “real languages”, like the many VLs. The idea of Stokoe and his successors was based on the search for common features between SL and VL. This approach made it possible to show the linguistic features of SL. However, it was limited to a very small number of signs with linguistic characteristics strictly similar to VL: they were *sign-words*, i.e. isolated signs which may be listed in SL vocabularies and easily translated with a single VL word. A further limit of these works was that they were focused on the manual components of SL, and this limited the exploration of the full expressive power of the language. Starting in the '80s, Christian Cuxac and his associates, after the implicit assumption that SL belongs to the “real languages”, developed a different model - the so-called “semiotic model” (Cuxac, 2000; Cuxac&Sallandre, 2007). Researchers along this line aimed at showing that SLis to be considered as a true language, with a richness and expressiveness comparable to VLs, though presenting characteristics fundamentally different from them. Differences are due to the use of the visual‑gestural channel rather than the acoustic‑vocal one. The works by Cuxac (1996, 2000) on the iconicity in LSF (French Sign Language) focused the attention on the non-manual components of signs, which are of paramount importance in many linguistic constructions of SL. Cuxac demonstrated that information in SL is transmitted also through eye gaze, facial expressions and body movements. Those researches then spurred analogous investigations on other SLs, e.g. the AntinoroPizzuto's research on Italian Sign Language (LIS) (AntinoroPizzuto, 2008; AntinoroPizzuto et al, 2010b; AntinoroPizzuto& Rossini, 2006). According to the semiotic approach, SL has two ways of expressing concepts, whichever the national variant (Cuxac&AntinoroPizzuto 2010) is:

- the "Lexematic Units" (LUs), i.e. the *sign‑words* analyzed by Stokoe, which can be easily translated using a single VL word; LUs are mostly characterized by manual production components (configuration, orientation, location and movement of the sole hands);
- the "Highly Iconic Structures" (HISs), i.e., complex structures with an underlying highly iconic content,which cannot be translated with a single VL word; in general, a HIS corresponds to several words, sometimes even to whole SL sentences (see Figure 1), and are produced by bringing into play both manual and non-manual components, e.g., hand position, facial expression, gaze direction, etc.

Some researchers still dispute the linguistic value of the HISs, considering it a mere *mimic gestural complement* to the LUs. However, it should be stressed that, while almost all the isolated signs in the SL dictionaries are LUs, several researches (including Cuxac&Sallandre, 2007; AntinoroPizzuto et al., 2008) have shown that 95% of the signs used in SL narratives are HISs (and not LUs), making it possible to firmly challenge any stance doubting their linguistic character (unless oneaccepts the existence of a language whose 95% is non-linguistic).

(a) 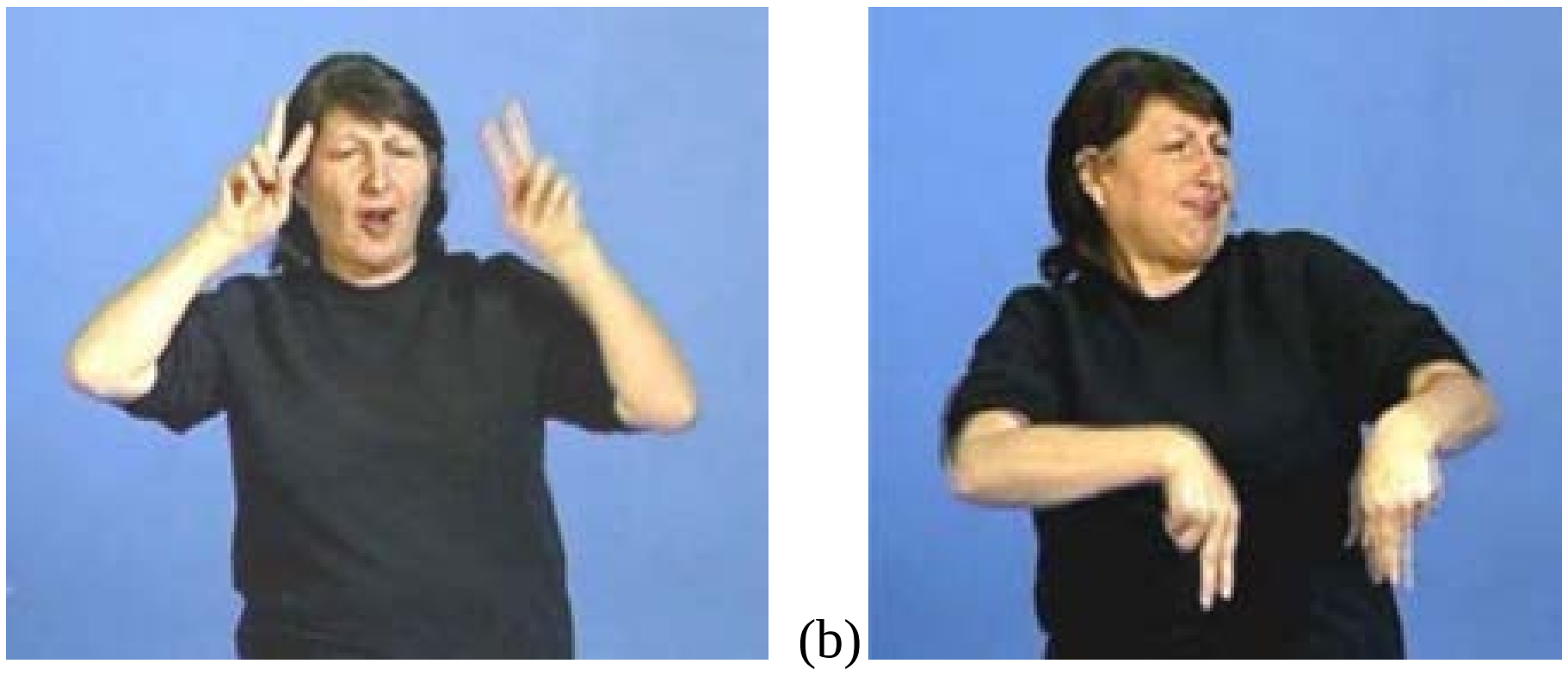 (b)

Figure 1 – French Sign Language from Corpus “LS Colin” (Cuxac et al., 2002): Lexematic Unit [HORSE] (a); Highly Iconic Structure of a horse galloping (b).[Garcia &Sallandre, 2014: fig. 11, p. 249 - used with permission]

The multiplicity of articulators that come into play in the production of HISs (the two hands, the body, the facial expression, the eyes, etc.) makes SL a language characterized by multi-linearity and simultaneity. In practice, VL produces 1D expressions that either develop along time (spoken), or along space (written). Differently from VL, SL expressions develop along 4 dimensions: 3D space and time. These characteristics make it impossible to carry out the “literal” translation *sign-to-word*from SLto VL. It should also be pointed out that even the syntactic structure of a national SL does not match that of the local VL: SL has a “visual” grammar. For instance, the sentence “John gives a flower to Mary” will be signed by placing the sign for John in a space A, the sign for Mary in a space B and making the sign for “to present” (which may assume a different manual form for a flower or a ball) move from A to B.

A final key feature of SL is that it is alanguage that has not developed a writing system accepted by a large part of the either national or worldwide Deaf Community. This trait is not rare, as only 10% of the world's languages (spoken, however, by 90% of the world's population – Lewis et al., 2015) have their own writing system. The fundamental difference with the other oral-only languages is that all the VLs can be represented by the International Phonetic Alphabet (IPA), which does proceed through the acoustic-vocal channel. This is impossible for SL, due to its nature, which is completely detached from such channel. In other words, the specificity of SLs does not allow an equivalent solution, due to the lack of a kind of atomic articulatory element with the same role of spoken letters. Furthermore, the characteristics of multilinearity, simultaneity and iconicity make it difficult, if not impossible, to adapted to the SL a system based on IPA. In the remaining part of this chapter, we will outline the different systems that, in spite of these difficulties, have tried to overcome the problem of the SL graphical representation. To describe these systems it is necessary to introduce and explain beforehand some linguistic terms, which are not usually used with the same meaning as in other fields, i.e. "transcription" and "writing".

- "Transcription" is the use of a graphics system to represent a linguistic phenomenon present in a spoken or signed production; IPA can be used to represent the phonology of a language, allowing to catch and analyze every detail of a phonological production; in the same way, a graphical system for SL should catch each movement that the signer does with the intent to convey a meaning; it is precisely to achieve this aim that the systems we comment afterwards have been developed.

- "Writing" has many valid meanings but the one we will take into account is the action which allows to generate a language production by means of a graphic representation of the language; contrary to the transcription, writing is a direct linguistic realization and not the graphical transposition of a previous production; none of the systems that we will examine, except SignWriting (Sutton, 1995; Bianchini 2012) ‑which underlies the software to be discussed ‑ was conceived to write the SL.

The lack of a writing system for SL is joined with the lack of a suitable transcription instrument. It must be emphasized that, for linguistic researchers, the graphical representation of SL is a real challenge, going well beyond a problem of accessibility: in fact, solving this problem would be the first step towards new frontiers for the scientific research on the language structures, with significant implications both for the “pure” linguistics and for pedagogy. Therefore, the main goal of the tools that we will discuss is to allow deaf users to find and use the language they prefer in interfaces as well as digital resources. In addition, a side effect of producing effective tools would be to facilitate linguistic research too.

**4. Examples of Strategiesto Represent/Communicate Digital Resources for the Deaf**

As discussed above, SLs have been long considered as a purely mimic form of communication. As a consequence, their full status of languages has been long argued as well, when not completely denied. This heavy historical heritage has long conditioned research and design in a twofold way: by hindering the comprehension of the real problems that deaf find with vocal languages, and by delaying the search for effective communication strategies to support them. This affects both the education process and, more recently, the access to the digital world of information. The lack of awareness of the deep structural (cognitive) differences between VLs and SLs has also nurtured the illusion to fill a purely sensorial gap by simply substituting auditory information by written text, as mentioned above by discussing WCAG guidelines. As already underlined before, the leitmotif of most accessibility guidelines that should address the needs of deaf people, is limited to textual captioning and audio content transcription (Fajardo et al., 2008). Concretely, this practice implies providing textual (VL) captioning wherever a resourceon the system(a website, an application, etc.), either audio or video clip, exploits the audio channel.It should be well evident at this point that this is not a solution, since both signing and non-signing people (those subject to oralist techniques) make similar mistakes with spoken and written language (AntinoroPizzuto et al., 2010a). On the other hand, despite the advances in the linguistic comprehension of their structure, SLs have not still developed a writing system of their own. This is the main reason for a special kind of digital divide involving deaf people, thatstill seriously limits the possibility to provide information, e.g., in electronic form and on the web, in a form “equivalent”, from the cognitive point of view, to the signed content. Think for example of the difference between a static text and a search: while the former can be substituted by a clip in SL, how is it possible to support the latter?

As a consequence, a deaf person finds in the digital world the same numerous and difficult barriers faced in everyday life. An example is given by the impossibility to exploit the “information scent” (Chi et al., 2001), while surfing through Internet. During a search, each user exploits “semantic” traces related to the desired information to judge which link to follow or which site to browse. The so called “trigger words”, i.e., words used for links and menu items, are a fundamental element in this process. From the point of view of deaf people, these semantic traces are generally available in a language that is not their “native” or preferred language, and in which they often have insufficient reading proficiency; we can experience a similar difficulty when browsing a site in a foreign language which we may know, but whose subtle hints are fully grasped only by a native speaker. As a matter of fact, some studies (Fajardo et al., 2008) have investigated and reported the difficulties that deaf users find in gathering the information they need through textual traces. The gradually increasing awareness of the research community about deaf difficulties is producing the first results. Accessibility issues have been mostly tackledin connection with more “serious” disabilities like blindness, but now they are taken into account also in relation with deafness. As examples of the new interest raised by related problems, we can mention the special issue ‘‘Emerging Technologies for Deaf Accessibility in the Information Society’’ of the international journal Universal Access in Information society, published in February 2008 (Efthimiou et al., 2008), and the frequent presence of related papers in Human Computer Interaction (HCI) conferences. Related research often deals with facial/gesture analysis for SL recognition from signed video (a review can be found in Ong and Ranganath, 2005), or with sign synthesis and sign language animation, possibly through avatars (Elliot et al., 2008).It is worth underlining that most research in automatic analysis of SL has focused on recognizing the lexical form of sign gestures in continuous signing, aiming at scaling well to large vocabularies. However, “successful recognition of lexical signs is not sufficient for a full understanding of sign language communication. Non-manual signals and grammatical processes which result in systematic variations in sign appearance are integral aspects of this communication but have received comparatively little attention in the literature.” (Ong and Ranganath, 2005). It is clear that this pessimistic observation refers to the goal pursued by most computer scientists, who approach this topic with the aim to build a kind of automatic interface, able to work as an interpreter on behalf of deaf people. However, most efforts along this direction have to face a number of technological as well as conceptual problems. A full direct translation between vocal languages and sign languages must overcome the problem discussed above of mapping a 1D (verbal) flow onto a 4D (signed in space and time) flow and vice-versa. Automatic interpretation of SLs, and especially a bi-directional one, is agoal which is still hard if not unfeasible to achieve. “For deaf persons to have ready access to information and communication technologies (ICTs), the latter must be usable in sign language (SL), i.e., include interlanguage interfaces. Such applications will be accepted by deaf users if they are reliable and respectful of SL specificities—use of space and iconicity as the structuring principles of the language. Before developing ICT applications, it is necessary to model these features, both to enable analysis of SL videos and to generate SL messages by means of signing avatars.” (Braffort and Dalle, 2008).Wecan conclude that automatic interpretation as well as “translation”is quite a Chimaera at present.

Less ambitious, but possibly more effective proposals in literature entail using recorded videos without any automatic interpretation attempt.Overall, despite the efforts made by the W3C (Caldwell et al., 2008), a widespread support for SL in the digital world is still far from being realized. Among the few exceptions, it is possible to mention the Dicta-Sign project (Efthimiou et al., 2010). Videos, more specifically signed videos, are typically the most widespread technique for SL inclusion. Concretely, signed videos are video clips representing one or more people producing signs. In this regard, some research projects have developed a series of techniques to allow deaf people to access digital information, through different forms of deaf-oriented hyperlinking. A seemingly simple idea is the basis of the Cogniweb project (Fajardo et al., 2008), which proposes two different techniques to equip a web page with SL videos intended to support navigation. In the first proposed technique, a video frame is located at the bottom of the page, and starts the corresponding SL sequence as the user moves the mouse over a link. In the second technique, a (mouseover-activated) signed video is included within each hyperlink. Two tests demonstrated that deaf people can navigate more efficiently using the second technique. Other, more advanced approaches aim at producing digital content using exclusively SL. The SignLinking system (Fels et al, 2006), for example, introduces an interaction modality featuring hyperlinks embedded within a video. They are defined asSignLinks. Each SignLink spans a time window within the video. An icon typically indicates the presence of a link. As the icon appears, the user can choose whether to follow the link or to keep watching the video.

The above, and similar current approaches, to make digital content accessible to the deaf appear effective and technologically smart and attractive. However, they share the same drawbacks. Even if many deaf communities have recently replaced many functions of writing by using signed videos, this is not always either possible or appropriate. As an example, some typical actions on the web arestill not possible and technically difficult to implement by video: to take notes, or to annotate a web resource (tagging), or to enter a query on a search engine. Actually, the first two mentioned actions could be (partially) supported by a multimodal annotation tool (see for example Bottoni*et al*., 2006). Nevertheless, as searching is involved, and most of all searching according to an intended "meaning", we find again all problems related to automatic interpretation and translation.Furthermore, videos are not anonymous: anyone can recognize the contributor simply looking at the video … unless wearing a mask. This holds many people back who would be otherwise eager to contribute. Finally, people cannot easily edit or add further content to a video that someone else has produced, so a wiki-like web site in SL is not possible (Efthimiou et al., 2010). In summary, videos lack the ease of handling and the variegate usability of a written expression. Also using signing avatars is not free from limitations. Automatic translation/transduction from text to a "convincing" signed sequences produced by an avatar is quite unfeasible, given the above discussion. Notwithstanding the many papers in literature (see Wolfe et al., 2015), there is still a long way to go (Kipp et al., 2011) even from the point of view of user acceptance. Moreover, carrying out the conversion in advance rises the same limitations of video, since it cannot be used for any normal real time activity on the web.

We can conclude that attempts for automatic interpretation as well as “translation” are at a very early and immature stage, and also using signed videos presents operational as well as accessibility limitations. As a consequence of these observations, the inclusion of written SL (and, as a consequence, the informatization of SL) rises to paramount importance in order to achieve an effective deaf-oriented accessibility design, and to ultimately mitigate the impact of the digital divide on deaf people (Borgia et al., 2012). Therefore, we embrace the project of supporting the ease of use of written forms of SL also in digital settings, to allow easy production and sharing of digital documents in a suitable notation. This would definitely improve accessibility for deaf users. In particular, our framework is based on the use of SignWriting, one of the proposed writing systems for SL.Some of these systems will be shortly presented in the next Section.

## 5. Writing systems for SL in literature

The method traditionally used by linguists to deal with languages is glossing. According to the main intended use, a *gloss* is a brief marginal notation of the meaning of a word or wording in a text. It may be produced either in the language of the text, or in the language of the reader, if they are different. Therefore, *glossing* is also writing one language in another, where the written information is the *gloss*. In the case of SL, it entails looking at someone signing and writing sign by sign,by further including various notations to account for the facial and body grammar that goes with the signs. In practice, there is no attempt for interpretation but rather for transcription. *Glosses* are labels in verbal language that are added to express the meaning of a SL expression in a very simplified way. Due to their popularity among linguists, they have long been a very common way of representing SLs, possibly adding annotations with further generally alphabetical symbols. The use of *glosses* isoften ratherconsidered a pseudo-notation (Bianchini, 2012), while true notation forms have been attempted along time. This is also reflected by the considerations by Wilcox and Wilcox (1997) regarding American Sign Language (ASL), that also apply to any SL: “Glosses are rough translations of ASL morphemes into English morphemes. The reader should understand, however, that the existence of glosses for ASL does not signify that ASL is English. The reader should also remember that glosses are not intended to be good translations into English of these ASL sentences.” Glossing SL means writing down a series of English words intended to represent signs (their “names”) in the order they would be produced in ASL. Words are annotated by additional information about the way each sign is made. Such information includes direction, kind of motion, possible repetition, and non-manual features, and is expressed by some standard symbols, or simply by describing the inflection used. An example is shown in Figure 2. A horizontal line placed over the gloss indicates simultaneous elements of ASL, while non-manual features are described above the line.

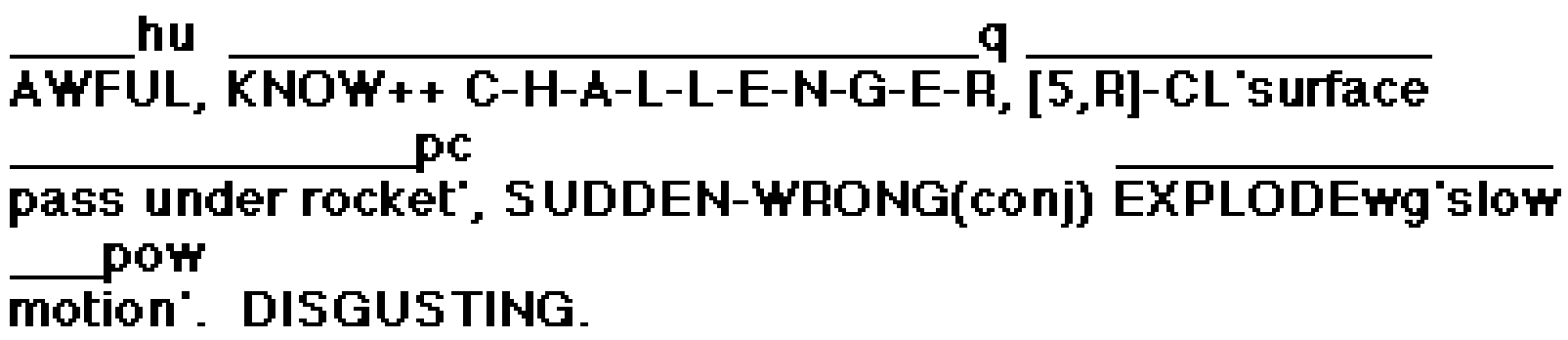

Figure 2 – Example of glossing for ASL

It should be understood that symbols used in glossing are rather related to the sentence construction, e.g., prosody, grammar, or repetitions, while the information on the form of the sign is lost (see AntinoroPizzuto et al., 2010b).

Devising a writing system especially dedicated to a language implies to understand its inner structure first. As reported above, one of the first researchers to appreciate the real linguistic and communication potential of SLs was William Stokoe's (1960). His seminal work still provides plenty of insight on the issue, so that is has been reprinted in 2005 (Stokoe, 2005). He gave a first fundamental contribution toward recognizing the richness and expressiveness of SLs, and one of the first notations used to "write" it. However, the limitations of such notation are twofold. First, it focuses only on the manual component of gestures. Stokoe's 4-parameter model is characterized by symbols that represent hand shape, hand orientation, location relative to other parts of the body, and movement. Second, it is much more appropriate for use in linguistic studies than in every day communication. This can be easily verified by looking at Figure 3.

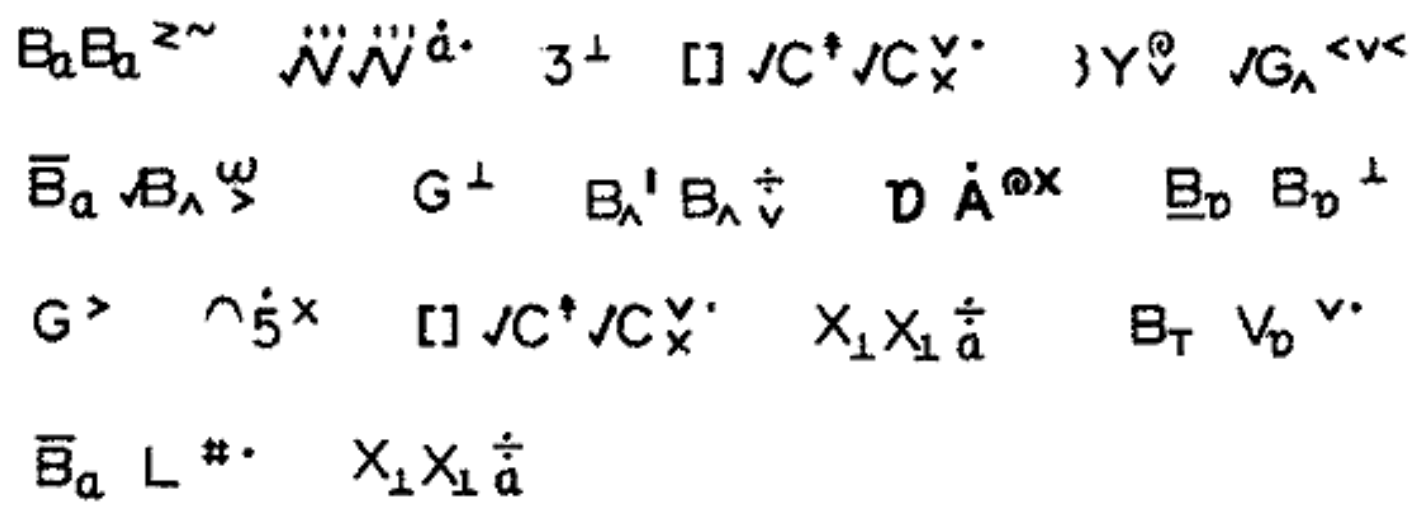

(a)

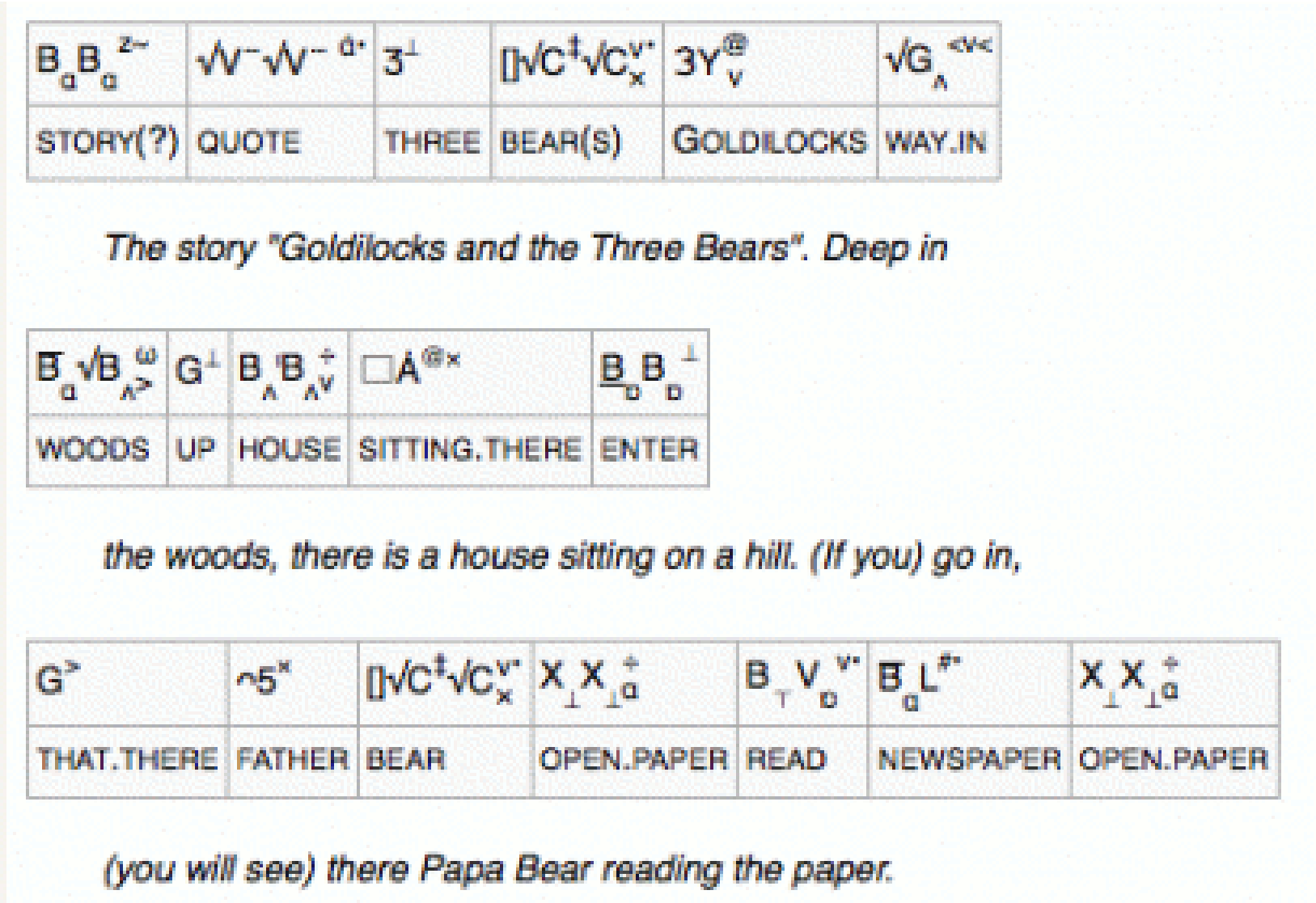

(b)

Figure 3 – A passage from Goldilocks in ASL transcribed in Stokoe notation (from Wikipedia)(a); glossing of the same passage (from http://scriptsource.org/)(b)

Last but not least, Stokoe notation basically relies on studies on ASL, and in this respect is not fully applicable without modifications for other SLs.

An attempt to better preserve the visual information contained in signs is the Hamburg Notation System (HamNoSys) (Hanke, 2004). It was born in 1985 from the initiative of a group of researchers at the University of Hamburg. It is based on Stokoe's 4-parameter model. Each sign (or word) is written  by assigning a value to each of these parameters. Its further revisions attemptto allow writing any signed language precisely, therefore overcoming some limitations of Stokoe notation. In addition, the writing system includes some support for non-manual features. The writing symbols are available as a unicode-based character set for various operating systems, and can be downloaded for free. The language is also a basis for a series of avatar controls. However, even HamNoSys was not devised for writing full sentences. Therefore, it continues to be popular for academic purposes and has undergone four revisions, but it is still unfeasible to use it as a script for everyday common communication. Figure 4 shows an example of the notation, that also exemplifies the lack of strict visual correspondence with the sign. It is not important to give the meaning of the sign, that actually might be different across SLs, but rather we can consider the scarce ability of the notation to immediately evoke the shape of the sign. To this regard, we can mention the distinction made by psychologists between two types of memory retrieval, according to the very well-known principle of recognizing vs. recall. Recognition refers to our ability to recognize something familiar, while recall entails the retrieval of related information from memory. The prevalence of the former over the latter positively influences learnability.

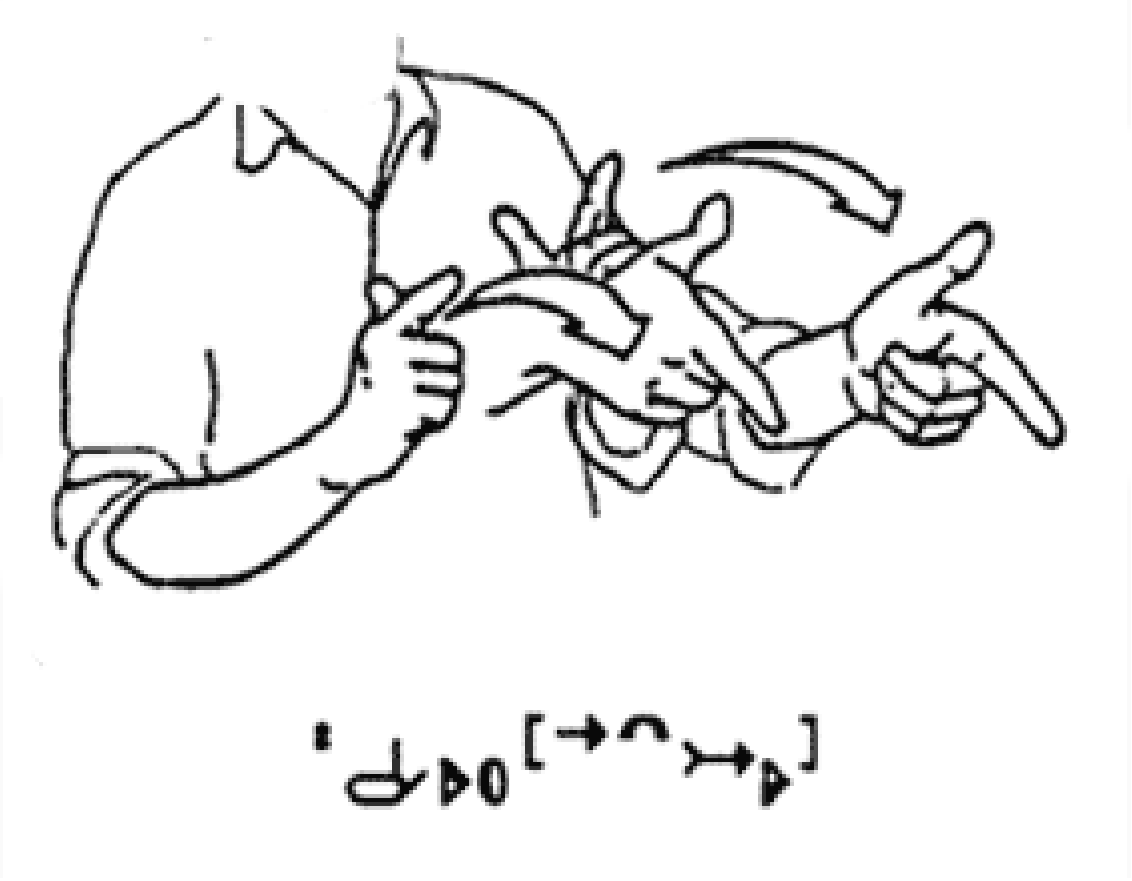

Figure 4 – A sign and its transcription in HamNoSys notation

While the basis for writing systems of VLs is the learned acoustic correspondence, the key for SL is to devise a visual correspondence, since the visual channel is exclusively used to communicate.Of course giving a full listing of possible notations for SL is out of the scope of this chapter. The interested reader can refer for example to the interesting web site [https://aslfont.github.io/Symbol-Font-For-ASL/.](https://aslfont.github.io/Symbol-Font-For-ASL/)

The above considerations leads us to consider one of the notations that are attracting a very wide interest in both Deaf community and among linguists carrying out research on SL, namely SignWriting.

Of course, the choice of a writing system is subordinated to the goal that one must pursue. For the point of view of the Human-Computer Interaction, SignWriting proves an extremely appropriate candidate to work with, since it has features that can rarely be found in modern SL writing systems. The system features a high level of iconicity, which in turn makes it very easy to learn and to master. Finally, the possibility to be employed in everyday use makes it the ideal candidate for a wide diffusion (Borgia et al., 2014).

SignWriting (Sutton, 1977, 1995) is an iconic writing system for SLs. In SignWriting, a combination of 2-dimensional symbols, called glyphs, is used to represent any possible sign in any SL. The glyphs are abstract images depicting positions or movements of hands, face and body. Figure 5 shows the Italian Sign Language (Lingua ItalianadeiSegni –LIS)  sign for Fun, written in SignWriting. Apart from the actual meaning of the sign, and from the use of a few conventions quite easy to grasp, it is immediate to notice the immediate evocation of the gestures used to produce the sign.

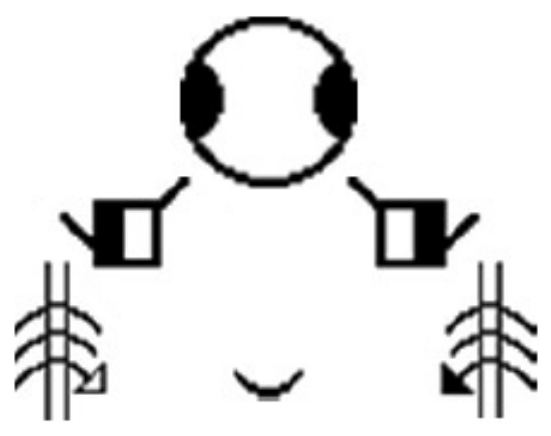

Figure 5: LIS sign for Fun, written in SignWriting.

The high iconicity of this system is due to the shapes of the glyphs themselves, that have been conceived to reproduce any movement or position of the upper part of the human body,in a stylized yet accurate way. SLs are characterized by the 3-dimensional spatial arrangement of gestures and by their temporal structure; in the very same way, the spatial arrangement of the glyphs in the page plays a core role. In fact, it does not follow a sequential order (like the letters of the written form of VLs), but the natural arrangement suggested by the human body.

Since SignWriting represents the actual physical formation of signs rather than their meaning, no phonemic or semantic analysis of a language is required to write it. A person who has learned the system can “feel out” an unfamiliar sign in the same way an English speaking person can “sound out” an unfamiliar word written in the Latin alphabet, without even needing to know what the sign means. Since 1996, the SignWriting standard also recommends to write following a vertical organization: writing signs in columns, one below the other. For this reason, most of the SignWriting texts available are written adopting this organization.

The set of movements and positions that a human body can produce from the waist up is huge. As a consequence, the set of glyphs that SignWriting provides to write down any sign is accordingly vast (about 30.000 units). The whole set of glyphs is referred to as the International SignWriting Alphabet (ISWA) (Slevinski, 2010a). The ISWA organizes the glyphs by dividing them into 7 categories, identified by following a very intuitive principle: each one covers a different anatomic part of the human body, with a small number of exceptions. Further distinctions, i.e. groups and base symbols, are present within each category. This helps keeping a logical and linguistic organization within categories,which would otherwise be too vast to manage. Categories, groups, base symbols and variations allow identifying a unique code (ISWA code) for each glyph within the ISWA. Such code is a key element for the digitalization of SignWriting, since it is much easier for a machine to work with 13-digit codes, rather than with raw unorganized symbols.

# 6. Digital SignWriting

## 6.1 Aim and Available Software

A pencil and a piece of paper are the only required items to produce signs using SignWriting. However, the system has already risen as an effective communication mean for deaf people in the digital world, thanks to a 30-year informatizationprocess started in 1986 by Richard Gleaves with the SignWriter computer program (Sutton, 1995).

First of all, there are bilingual websites or blogs (e.g. Frost, 2006) accessible from both hearing and deaf users by supporting both VL - in English - and American Sign Language (ASL). Furthermore, unlike signed videos, the inclusion of written SL also enables the creation of wiki-like websites. In fact, an ASL Wikipedia Project (Sutton, 2012) is currently underway. The ASL articles in the ASL Wikipedia are written by deaf users, mainly as ASL translations of VL articles in the English Wikipedia. The goal of the project is to provide a bilingual, educational, informational tool intended to “provide information about the world to deaf and hearing people who use ASL as their daily, primary language” (Sutton, 2012).

Dictionaries and sign databases are among the online resources available in SignWriting. Such repositories, such as SignBank (Sutton, 2010), provide SL users with an archive of illustrations, sign language video-clips and animations. Most importantly, SignBankprovides a way to find words by looking up signs, using SignWriting symbols or VL keywords. This kind of digital artifact may prove a valuable asset for those who use Sign Language on a daily basis (especially for content authoring), whether they be deaf or hearing people.

Finally, most SignWriting digital resources are mainly available only thanks to a specific class of software, i.e. SignWriting digital editors. Such applications are the tools that enable the creation of digital signs written in SignWriting. In other words, they are critical for the informatization of SignWriting. Many applications have been produced by different research teams. Sutton's official SignMaker editor (Slevinski 2010b) is one of the most popular, but a fair number of alternatives are available, such as SWift, DELEGs and SignWriter studio. Most SignWriting digital editors basically provide the same functionalities. Despite differences in interface design and implementation existing from one editor to another, such functionalities are:

- Search for one or more glyphs belonging to the ISWA;
- Insert the chosen glyph(s) onto an area designated for the composition of the sign;
- Save the sign in one (or more) formats.

In general, the informatization of a writing system for SL poses different challenges. First of all, computer scientists need to devise effective and efficient ways of dealing with a set of symbols as large as the set of movements and positions that can be produced from the waist up. When designing a SignWriting digital editor, for instance, the large cardinality of the ISWA set might become a major problem for the application. If addressed incorrectly, it might affect both logic and presentation layers. It is necessary to get as close as possible to the "aureamediocritas" between the unrestricted access to the data (the glyphs) and the presentation of a human-manageable amount of information. Furthermore, the rules underlying the composition and the organization of a SL writing system are (generally) very different for those holding for a VL system. For instance, SignWriting grants a very high degree of freedom to its users. There is no high-level rule at all about the composition itself: no restriction is set to the number of glyphs within a sign, to their possible spatial arrangement and to their relative positioning.

### 6.2 A Digital Editor for SignWriting: SWift

In this section, we introduce SWift (SignWriting improved fast transcriber) (Bianchini et al., 2012b), an editor conceived by a research team of (both deaf and hearing) linguists, SignWriting users, and computer scientists in the field of HCI. SignWriting editors are seldom designed holding usability as the main focus. In most cases the main goalsare giving the user an unrestricted access to the glyphs, and providing the necessary functionalities to manage the created sign. On the contrary, SWift has been designed in close collaboration with a true sample of its target users, namely the hearing and deaf researchers of CNR ISTC (Istituto di scienze e tecnologiedellacognizione– Institute for science and technologies of cognition), in full compliance with the core principles of User-Centered Design (Norman & Draper, 1986) and Contextual Design (Wixon, et al., 1990). Both design techniques require to involve final users as the main stakeholders in the design process. The first technique requires giving extensive attention to needs, preferences and limitations of the end users of a product. This must be done at each stage of the design process. The second technique requires that researchers aggregate data from customers in the very context where they will use the product, and that apply their findings during the design stages. Actually, a number of further stakeholders should be involved when designing for people with special needs (De Marsico et al., 2006). These include: experts in specific disabilities, that can suggest the best ergonomic strategies to support these users; experts in the target application domain, that can suggest alternative channels and/or message modifications to better reach these users; and experts in the accessibility guidelines regarding users with special needs. It could prove very helpful to follow some specific requirements, while designing the User Interface (UI) of an accessible application for deaf people, e.g. a digital editor like Swift. They are listed below.

- Intuitiveness: the user shall be relieved from the burden of learning the UI, whereas the UI can simply be understood; with this purpose in mind, each function shall be presented (and work) in a intuitive and familiar way.
- Minimization of information: each screen shall present a small amount of essential information, in order to avoid overwhelming the user with a cluttered UI.
- Look and feel: icons shall be simple, large and familiar; if their meaning remains unclear, mouse-over-triggered animations/videos could be embedded in the buttons/links to guide the user; dealing with text labels in the UI might be difficult for deaf people, therefore such elements shall be kept at a minimum (Perfetti&Sandak, 2000).
- User-driven interface testing: each (ot the most important at least) change in the UI shall be discussed and tested with a team including the target users of the application; the testing shall involve high-level aspects as well as low-level ones (such as the spatial placement of buttons within the UI, etc.).

The UI of SWift (shown in Figure 6) is an example of deaf-accessible UI, which has been designed meeting all of the above-mentioned requirements. Like other digital editors, SWift provides an area to compose the sign. Such component is referred to as the *sign display*; it is a whiteboard-resembling area whose purpose is to show the sign that the user is currently composing. A sign is composed by dragging a number of *glyphs* from the *glyph menu* and dropping them on the *sign display*. Once they are placed there, they become both draggable (to be relocated at will) and selectable (for editing purposes). The far larger size of the *glyph menu* with respect to the *sign display* may seem strange, but the reason for this will be clear soon.

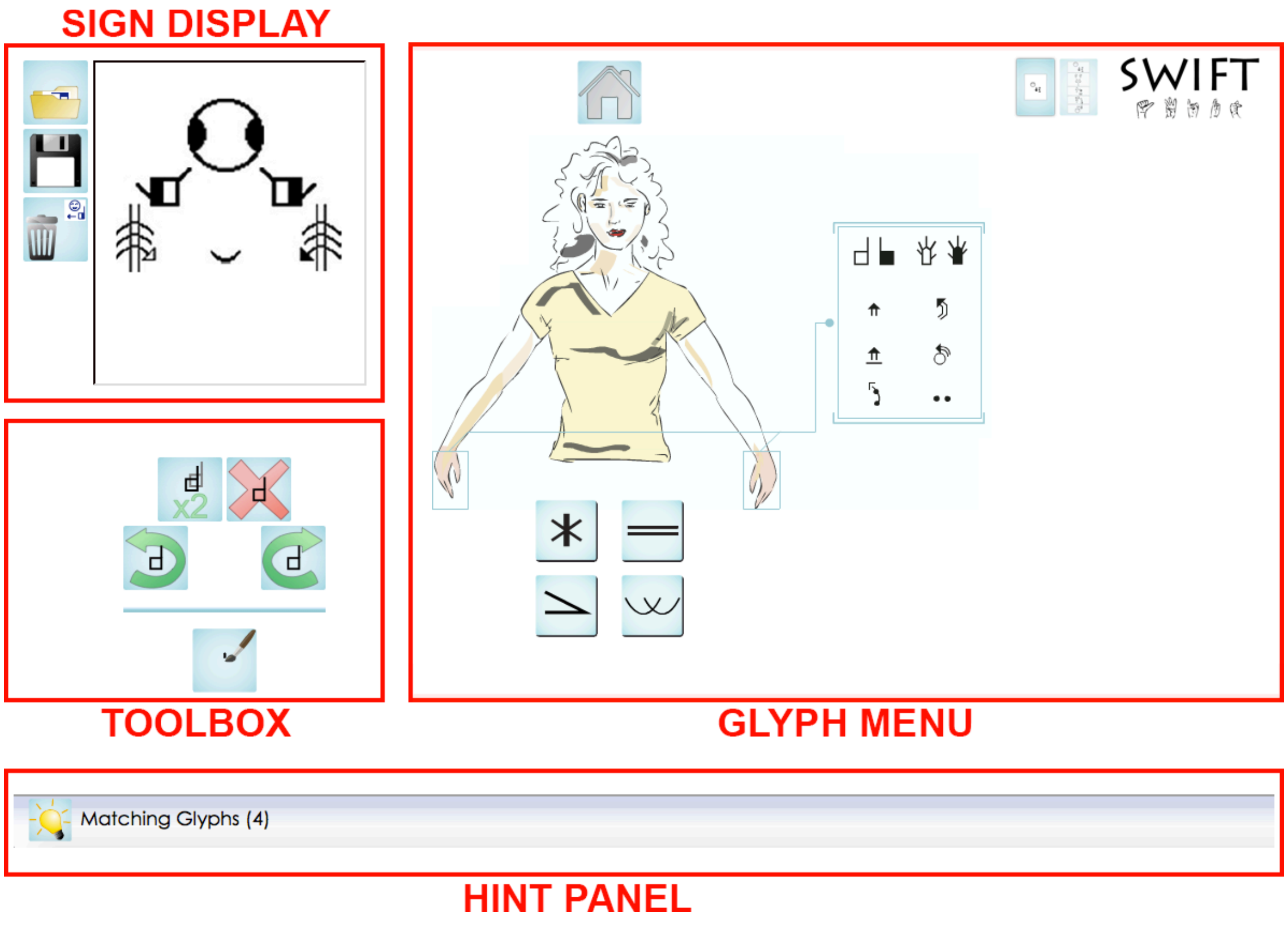

Figure 6: Home screen of SWift, divided into 4 functional areas.

The *glyph menu* allows the user to search any glyph within the ISWA. Once the user finds the desired glyph, he/she can drag it and drop it on the *sign display*, in order to include it within the sign that is under construction. Most efforts were devoted to make the interaction with the *glyph menu* fast and effective. The underlying concept is basically “Why browse, when you can search?”. For this reason, the *glyph menu* features a search system which implements a completely different approach with respect to its competitors (Bianchini et al., 2012a). In order to support SignWriting beginners during the composition, the *glyph menu* has been designed to present a stylized human figure (Figure 6), folksily called *puppet*, as a starting point for the search of a glyph. The purpose of the *puppet* is making the search of a glyph easier and faster, by making the choice of anatomic areas (categories and groups) more straightforward. In other words, it enforces recognition vs. recall. The buttons below the *puppet* represent group of glyphs related to signing aspects apart from body parts, e.g., repetitions and contact points. By choosing an anatomic area of the *puppet*, or one of the buttons below, the user accesses the dedicated search menu for that area or for that kind of item. After the user clicks, the *puppet* and the buttons beneath are reduced and shifted to the left, but remain clickable and form a navigation menu together with the button to return to the *glyph menu*'s home screen. This allows freely navigating from one area to another. A red square around the selected area reminds the user’s choice, like breadcrumbs. In the central part of the menu, a label and an icon show what kind of glyphs are available using the group of boxes beneath. These boxes are referred to as *choice boxes* and guide the user during the search for a glyph. They display groups of options: the user can choose only one element from each of them, in any order, and each choice progressively restricts the set of symbols displayed as response to the user search. Figure 7 shows an example. The configuration of the *glyph menu* at any step explains the need for a large area in the interface. Once a glyph is found, it can be directly dragged and dropped on the *sign display*.

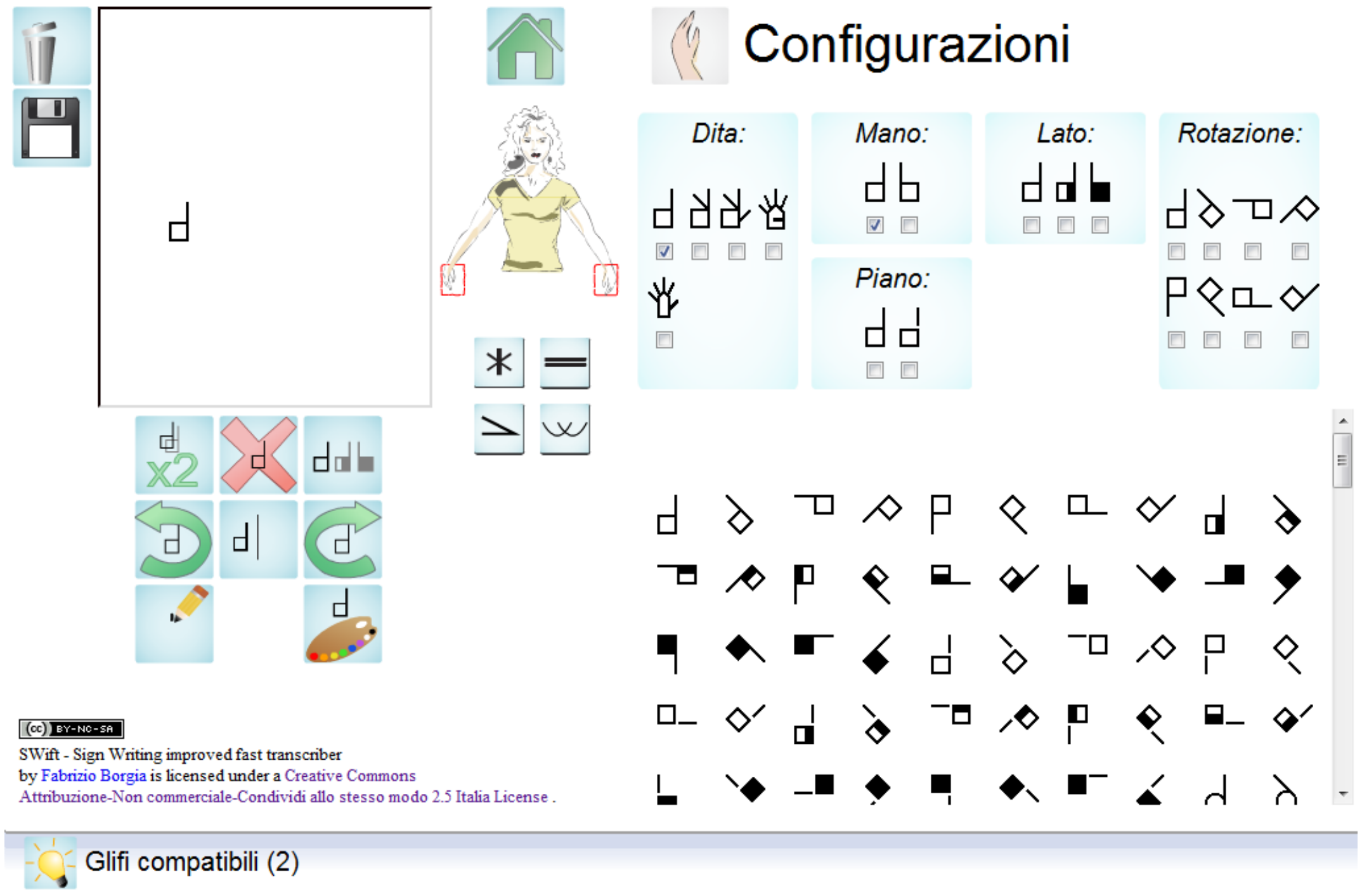

Figure 7: An example of navigation support.

Finally, the *hint panel* is one of the innovations that distinguishesSWift from other digital editors, since it implements a prototype predictive sign composition for SL. Many studies, such as Curran, et al. (2006), demonstrate that predictive text is an important aid to communication when handling the set of characters of a VL, which are in the order of magnitude of tens. It is easy to realize how this can improve the interaction with a set of tens of thousands symbols like ISWA. The *hint panel* enables predictive sign by showing in real-time, as the composition process is underway, a set of glyphs which are *compatible* with those the user already entered in the *sign display*. Compatibility is computed and updated according to the rate of co-occurrence of the glyphs in a database of known and already stored signs. The glyphs suggested in the *hint panel* are immediately available to be inserted into the *sign display*. With such action, the user can save the effort and the time required to search for each glyph from scratch.

### 6.3 Assessing Usability of Applications for Deaf Users

Assessing the reliability and the usability of a deaf-accessible application is of paramount importance during the development lifecycle. It is advisable, yet not always possible, to conduct test sessions with participants representing the actual end users of the application. In the case of a deaf-accessible application, deaf people should be included within the group of the participants; in the particular case of a digital editor (such as SWift) it was necessary to test the application with deaf people being also proficient in SignWriting. However, whenever very specific requirements have to be met to reflect the real target community, a statistical sample in the order of tens is often the best that one can achieve. For instance, this was the case with SWift.

The choice of the proper set of tools is very important during the usability assessment of an application. Most tools, however, do not consider the possibility that the sample is composed of people with special needs, such as deaf people. Currently, a very restricted set of tools and studies is available to this regard. One of the most popular and effective tools for usability testing is the Think-Aloud Protocol (TAP) (Lewis &Rieman, 1993). However, since deaf participants cannot actually “Think Aloud”, different research teams around the world adapted TAP in order to include SL. The first adaptation was suggested by Roberts and Fels (Roberts &Fels, 2006), who performed a number of tests with deaf users adopting a TAP-based protocol. Along the same line, an adaptation of the TAP, called Think by Signs (Bianchini et al., 2012a) is described here. The Think by Signs is a bilingual (VL and SL) test methodology, so it can be employed with both hearing and deaf people. The TAP itself partly interrupts the attention flow of the user, since it engages cognitive resources along a different track. However, the positive aspect is the possibility to express one’s own impressions in real-time, without the possible bias due to the final outcome. In the specific case of deaf people, the interruption of the flow of the attention is even more concrete: the participant will inevitably stop using keyboard and/or mouse, since one or two hands are required in order to produce signs. However, signing during different actions is typical of the way deaf people have to communicate while performing a task, so this does not affect the outcome of the test.

More specifically, the Think by Signs test is composed by two moments. The *welcome time*starts as soon as the participant sits down in front of the computer. The system displays a welcome screen containing a signed video (on the left part of the screen) and its VL translation (on the right part). Consistent with the rules of the TAP, the greeting sequence contains: a brief thanks for the participation, an explanation of the structure and the rules of the test, and finally a reminder about the purpose of the test (which is not conceived to test the skills of the participant, but rather to test the capabilities and the usability of the software). The latter should help the participant in feeling at ease with the test. The *test time* follows the welcome time, and it represents the core part of the procedure. The participant is required to perform a list of tasks to test the functions of SWift and their usability. As stated by the rules of the TAP, during this phase the participant is asked to sign anything that comes to his/her mind. Given the possible high number of tasks, it is appropriate to alternate, when possible, simple assignments with complex ones, to avoid tiring the participant. For the same reason, the participant might need a reminder about the task that is currently underway. To address such need, it is recommended to design a task list to guide the participant. Several options are available to create a bilingual task list. In the first place, the test designer should decide whether to delegate the SL inclusion to an electronic device (with signed videos illustrating each task) or to an interpreter. In the second place, consistently with the first choice, the specific role and responsibilities of the device/interpreter must be clearly defined. To this regard, (Borgia, 2015) observed that the involvement of an interpreter usually makes the user feel more comfortable than a recorded explanation. This is also due to the possibility to indirectly or directly ask question about the required activities, and, of course, increases the probability of a correct understanding of the tasks too. For this reasons it is advisable to choose an option involving an interpreter. Given this, it is further necessary to define the way the interpreter has to act during the test. One of the best options is to have the interpreter always provide an initial SL translation of the task. Therefore, at the beginning of each task, a card is presented to the participant. The card contains a simple, direct VL question which identifies the task. As the card is presented, the interpreter signs the question to the participant. Afterwards, the participant is allowed to ask questions to the investigator, using the interpreter as intermediary. The answer may follow only if it does not affect the outcome of the test. Last but not least, deaf people disclose their attitudes and sensations through a very high variability of SL non-manual components. Due to this characteristic, it is of paramount importance to capture the user's face when recording a test session. Different spatial settings for the test room are available in (Roberts &Fels, 2006) and (Borgia, 2015).

In compliance with the guidelines discussed above, the sample of participants to the usability assessment of SWift was mainly composed by deaf people, and a Think by Signs test was carried out. At the end of the test, the participants were also asked to fill in a questionnaire to assess their subjective satisfaction with specific aspects of the UI. The tool chosen for the job was the Questionnaire for User Interaction Satisfaction (QUIS) (Chin et al., 1988). Like the TAP, even the QUIS was adapted to fit in a bilingual test environment, by presenting any suitable question among the set of original ones (and any possible answer) both in VL and SL with the aid of signed videos.

### 6.4 Optical Glyph Recognition for SignWriting: SW-OGR

Despite the efforts of different research teams, SignWriting digital editors are still far from granting the user an interface which is able to emulate the simplicity of handwriting. Actually, any software solution developed to support the production of SignWriting documents relies heavily on Windows Icons Menus Pointer (WIMP) interfaces, both for accessing the application features and for the sign composition process itself. The problem is all but a theoretical one. Even dealing with word processors (Latin alphabet), the users often feel the higher complexity and the slower composition time with respect to handwritten text production. Given its huge number of glyphs, this especially holds for people using SignWriting, in particular for deaf people. In fact, they are far more accurate, fast, and comfortable using the paper-pencil approach rather than dealing with the (more or less) complex interaction styles of a digital editor. For this reason, the design of a new generation of SignWriting editors has been planned (Borgia et al., 2014), able to relieve the user of any, or most at least, burden related to clicking, dragging, searching, browsing on the UI during the composition process of a sign. The purpose of the designers is to implement an interaction style which is as similar as possible to the paper-pencil approach that humans normally use when writing or drawing. In order to achieve such goal, it is necessary to integrate the digital editor with another software module (see Figure 8), more specifically an Optical Character Recognition (OCR) engine. The purpose of this additional module purpose is to operate the electronic conversion of images, containing handwritten or printed SignWriting symbols, into ISWA-encoded SignWriting texts. Such technique is known as the SignWriting Optical Glyph Recognition (SW-OGR) (Borgia, 2015).

Of course, since WIMP is currently the easiest, most common interface style in the world, it cannot be totally left behind, because it is necessary to access the features of most applications. Nevertheless, dismissing the WIMP style during the sign composition, which is the core part of any SignWriting editor, could prove a rewarding choice.

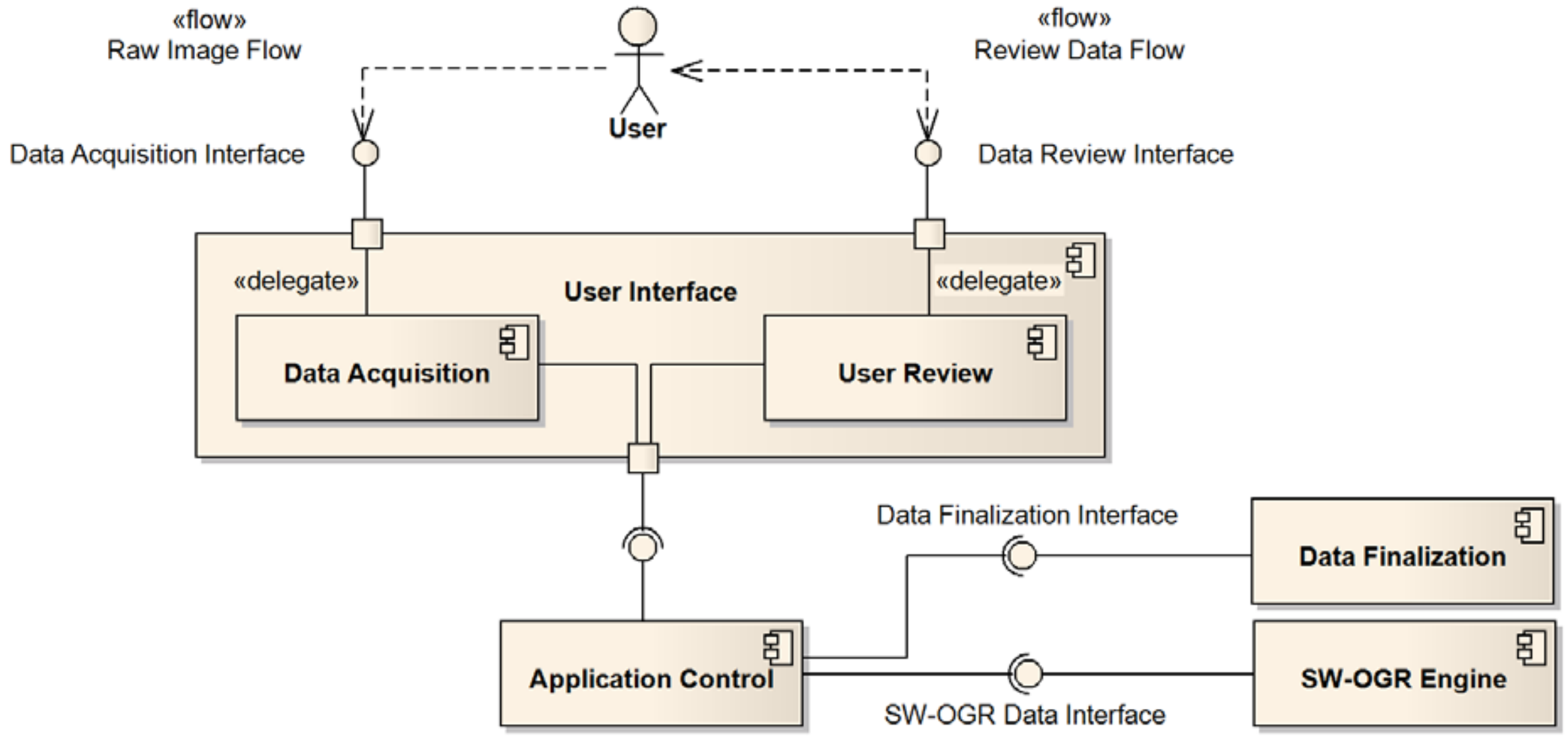

Figure 8: Component diagram for a new generation of SignWriting editors featuring a SW-OGR Engine.

A conceptual schema of the new generation of OGR-powered SignWriting digital editors is illustrated by the diagram in Figure 8; the application is composed by:

- User Interface, which includes two sub-components.
  - Data Acquisition, which provides the user with a simple interaction style to compose signs, focusing on intuitiveness (or, better, transparency ) and accuracy; this component is also designed to collect the data produced by the user (typically an image) and to pass it to the control component.

○ User Review, which comes into play after the recognition; it allows the user to make corrections (no recognition is 100% accurate) and/or add further data.

● Application Control, which implements the (Model-View-Controller pattern) controller of the application; among other tasks, this component also coordinates the data flow between the UI and the SW-OGR Engine.

● SW-OGR Engine, which is the model component of the application; its purpose is to provide a fast and accurate recognition of all (or most) glyphs handwritten or submitted by the user; the final product of the recognition is an image and an associated data file containing the ISWA codes (and the coordinates) of the recognized glyphs.

● Data Finalization Module, which saves the user-reviewed data in the requested format (image file, XML file, database entries, etc.).

The SW-OGR engine performs the recognition of SignWriting texts by only working with the geometric and topological features of the symbols, and with their topological relationships. The recognition also takes advantage of context-dependent information, such as the knowledge of the structure of the ISWA, i.e. its categories, groups, etc. The engine is intended to serve a twofold purpose: first of all, it can be embedded within existing SignWriting editors, such as SWift, in order to provide a prompt support for handwriting, and make the composition process much faster and comfortable for everyday use. In addition, it is worth noting that a considerable number of paper-handwritten SignWriting corpora exists, gathered from different communities around the world. Those corpora are an invaluable asset, and they could become even more useful if digitalized.

**7. Conclusions and Integration Perspectives**

In the context of deaf oriented accessibility, applications to support written SL, whether they are digital editors or highly specialized applications (like the SW-OGR engine), are not intended to be separated realities. In the specific case of SignWriting, the integration perspectives of SignWriting digital editors and SW-OGR are very interesting. In fact, observing the diagram in Figure 8, it is possible to infer that the application handling the UI (and the Application Control) could be one of the already existing digital editors, since the required features and application interfaces substantially correspond. Since any application concerning digital SignWriting shares the same way of representing signs, the integration is easy to implement. In fact, both digital editors and SW-OGR represent a sign as an XML document using the SignWriting Markup Language dialect (SWML). Within the document, each sign is associated to the list of its component glyphs, storing their ISWA codes and their spatial coordinates in the sign space (like the Sign Display in Figure 6). As a consequence, a document produced by SW-OGR engine can be read and updated using SWift. Such interoperability is the idea underlying a multi-application framework, whose purpose is making SignWriting effectively exploitable as a communication mean and as a learning support for deaf people. Such framework, named SignWriting-oriented resources for the deaf (SWord), has already been sketched (Borgia et al., 2014), and is under development. SWord is intended to support the acquisition of a corpus of signs from 2 possible sources: user-composed via digital editors (SWift, in particular), and digitized SignWriting corpora (currently on paper) via SW-OGR. An intermediate goal of the framework is to use these acquisition methods to gather a significant amount of signs, to be stored in electronic form together with their decomposition into glyphs. A set of signs of this kind is referred to as *structured corpus* (Borgia, 2015). The purpose of a corpus prepared in this way is to allow the identification of recurring patterns in the composition of the signs, and the computation of relevant statistics on the transcribed form of the signs. These elements are of paramount importance in order to gain a deeper understanding of the rules of SLs. Ultimately, the precise linguistic and production information stored by each SignWriting glyph can allows computer scientists to use them to determine the movements and expression of a signing avatar in a very accurate and satisfactory way. A similar approach has been already adopted by Karpov and Ronzhin (2014), by implementing a 3D signing avatar for Russian sign language, which is based on signs represented with HamNoSys. In this way, one might avoid using written text, which is almost impossible to automatically translate in SL, to derive the avatar behaviour. Using the intermediate form of SignWriting as an alternative starting point to guide the avatar, people using Sign Language may be supported in a number of activities that would otherwise require the use of a VL (e.g. e-learning), even without directly knowing SignWriting.

Last but not least, “transduction” of gestures from signed videos into SignWriting documents is the final step in the plan for the overall SWord. This step is at present at a very early stage, due to difficulties underlying computer vision-based approaches to the problem. These difficulties are mostly due to occlusion and self-occlusion of relevant body parts during signing, and to the “tokenization” of a visual sequence. However we are strongly convinced that it is worth devoting more efforts to this and similar projects, to provide full expressive possibilities in the digital world even to deaf people.

**8. In Memory**

We would like to remember Elena AntinoroPizzuto and Patrice Dalle, that with their great enthusiasm,deep competence and scientific sensibility spurred research as well as debate on deaf world and on its fascinating language.

**9. Acknowledgements**

We thank the group of hearing and deaf researchers of CNR ISTC (Istituto di scienze e tecnologiedellacognizione– Institute for science and technologies of cognition), since working with them lead us to a deeper comprehension of deaf problems and sensibility. Thanks to their collaboration our framework is taking shape and substance

**10. References**

Antinoro Pizzuto,E., Rossini P, Sallandre M.-A.,& Wilkinson E. (2008). *Deixis, anaphora and Highly Iconic Structures: cross-linguistic evidence on American (ASL), French (LSF) and Italian (LIS) Signed Languages*. *In* M. de Quadros (ed.) “Sign Languages: spinning and unraveling the past, present and future”. Editora Arara Azul.Petrópolis/RJ. Brazil: 475‑495.

Antinoro Pizzuto, E.,&Rossini, P. (2006). Deixis, anafora and highly iconic structures: cross-linguistic evidence on American (ASL), French (LSF) and Italian (LIS) signed languages. In Proccedings of TISLR 9 International Congress "Theoretical issues in Sign Language research" (Florianopolis, 6/12/2006).

Antinoro Pizzuto, E. (2008). Meccanismi di coesione testuale e Strutture di Grande Iconicità nella Lingua dei Segni italiana (LIS) e altre Lingue dei Segni. In Atti del Convegno nazionale "La grammatica della Lingua Italiana dei Segni" (Venezia, 16/05/2007). Venezia, IT: Cafoscarina.

Antinoro Pizzuto, E., Bianchini, C.S., Capuano, D., Gianfreda, G., &Rossini, P. (2010a). Language resources and visual communication in a deaf centered multi-modal e-learning environment: issues to be addressed., in: Proceedings of the LREC 2010.

Antinoro Pizzuto, E., Chiari, I., &Rossini, P. (2010b). *Representing signed languages: theoretical, methodological and practical issues*. In M. Pettorino, F.A. Giannini, I. Chiari & F. Dovetto (eds.), Spokencommunication. Newcastle upon Tyne: Cambridge Scholars Publishing, pp. 205-241.

Bianchini, C.S., Borgia, F., &De Marsico, M. (2012b).SWift A SignWriting Editor to Bridge between Deaf World and E-learning.In 2013 IEEE 13th International

Conference on Advanced Learning Technologies (Vol. 0, p. 526-530).Los Alamitos, CA: IEEE Computer Society.

Bianchini, C.S., Borgia, F., Bottoni, P.,&De Marsico, M. (2012a). SWift: a SignWriting improved fast transcriber. In G. Tortora, S. Levialdi, & M. Tucci (Eds.), Proceedings of the International Working Conference on Advanced Visual Interfaces (p. 390-393). New York, NY: ACM. doi:http://doi.acm.org/10.1145/2254556.2254631

Bianchini, C.S. (2012). *Analysemétalinguistique de l'émergence d'un systèmed'écriture des Langues des Signes : SignWriting et son application à la Langue des SignesItalienne (LIS)*. PhD dissertation, University Paris VIII: 512 p.

Borgia, F. (2015). *Informatisationd'uneformegraphique des Languesdes Signes: application au systèmed'écriture SignWriting*.PhDdissertation. Universite Toulouse III - Paul Sabatier, Toulouse, France.

Borgia, F., Bianchini, C. S., Dalle, P., & De Marsico, M. (2012). Resource production of written forms of Sign Languages by a user-centered editor, SWift (SignWriting improved fast transcriber).In International Conference on Language Resources and Evaluation - LREC (pp. 3779-3784).

Borgia, F., Bianchini, C., &De Marsico, M. (2014). Towards Improving the e-learning Experience for Deaf Students: e-LUX. In C. Stephanidis& M. Antona (Eds.), Lecture Notes in Computer Science: Universal Access in Human-Computer Interaction (Vol. 8514, p. 221-232). Berlin, Germany: Springer. doi: 10.1007/978-3-319-07440-5 21

Bottoni, P., Levialdi, S., Pambuffetti, N., Panizzi, E., &Trinchese, R. (2006). Storing and retrieving multimedia web notes.*International Journal of Computational Science and Engineering*,*2*(5-6), 341-358.

Bottoni, P., Borgia, F., Buccarella, D., Capuano, D., De Marsico, M., & Labella, A. (2013). Stories and signs in an e-learning environment for deaf people. *Universal access in the information society,* 12(4), 369-386.

Braffort, A., &Dalle, P. (2008). Sign language applications: preliminary modeling. *Universal access in the information society,* 6(4), 393-404.

Caldwell B., Cooper M., Guarino Reid L.,&Vanderheiden G. (2008). *Web Content Accessibility Guidelines (WCAG) 2.0.*World Wide Web Consortium. [http://www.w3.org/TR/WCAG20/ accessed 05/01/2015]

Chi, E. H., Pirolli, P., Chen, K., &Pitkow, J. (2001, March). Using information scent to model user information needs and actions and the Web. In Proceedings of the SIGCHI conference on Human factors in computing systems (pp. 490-497). ACM.

Chin, J. P., Diehl, V. A., & Norman, K. L. (1988). Development of an instrument measuring user satisfaction of the human-computer interface. In Proceedings of the SIGCHI conference on Human factors in computing systems (pp. 213-218). ACM.

Curran, K., Woods, D., &O'Riordan, B. (2006). Investigating text input methods for mobile phones. *Telematics and Informatics* , 23 (1), 1–21.

Cuxac C. &AntinoroPizzuto E. (2010). *Émergence, normeet variation dans les langues des signes: versuneredéfinitionnotionnelle.In*: B. Garcia, M. Derycke (eds) "Sourdsetlangues des signes: norme et variations". LangageetSociété, 131: 37‑53.

Cuxac C. &Sallandre M-A.(2007). *Iconicity and arbitrariness in French Sign Language : Highly Iconic Structures, degenerated iconicity and diagrammatic iconicity. In* E. AntinoroPizzuto, P. Pietrandrea& R. Simone (eds.): Verbal and Signed Languages: Comparing structures, constructs and methodologies. Mouton de Gruyter, Berlin: 13-33.

Cuxac, C. (2000). *La Langue des SignesFrançaise (LSF): les voies de l'iconicité*. Ophrys, Paris.

Cuxac C., Braffort A., Choisier A., Collet C., Dalle P., Fusselier I., Jirou G., LeJeune F., Lanseigne B., Monteillard N., Risler A. &Sallandre M.-A. (2002). Corpus LS-COLIN. [http://cocoon.huma-num.fr/exist/crdo/meta/crdo-FSL-CUC021_SOUND]

Cuxac, C. (1996). *Fonctionset structures de l'iconicité des langues des signes*. Thèse d'État, Université Paris V.

De Marsico, M., Kimani, S., Mirabella, V., Norman, K. L., & Catarci, T. (2006). A proposal toward the development of accessible e-learning content by human involvement. *Universal Access in the Information Society*, 5(2), 150-169.

Efthimiou, E., Fotinea, E., Glauert, J. (eds.) (2008). Special issue: "Emerging Technologies for Deaf Accessibility in the Information Society". *Universal Access in the Information Society* February 2008, Volume 6, Issue 4.

Efthimiou, E., Fotinea, S., Hanke, T., Glauert, J., Bowden, R., Braffort, A., Collet C., Maragos, P., &Goudenove, F. (2010). DICTA-SIGN: Sign Language Recognition, Generation and Modelling with application in Deaf Communication. In Proceedings of the 4th workshop on the representation and processing of sign languages: Corpora and sign language technologies (CSLT 2010), 2010, satellite workshop of the LREC-2010 conference (p. 80-83). Valetta, Malta.

Elliott, R., Glauert, J. R., Kennaway, J. R., Marshall, I., & Safar, E. (2008). Linguistic modelling and language-processing technologies for Avatar-based sign language presentation.*Universal Access in the Information Society*, 6(4), 375-391.

Fajardo, I., Arfe, B., Benedetti, P. &Altoé, G.M. (2008).Hyper- link format, categorization abilities and memory span as contributors to deaf users hypertext access. *Journal of Deaf Studies and Deaf Education*, 13(2), pp. 241–256.

Fajardo, I., Parra, E., Cañas, J., Abascal, J., & Lopez, J. (2008, July). Web information search in Sign Language. Paper presented at Informal Learning on the Web: Individual Differences and Evaluation Processes (Symposium), XXIX International Congress of Psychology, Berlin, Germany.

Fajardo, I., Vigo, M., Salmerón, L. (2009). Technology for supporting web information search and learning in Sign Language.*Interacting with Computers*, 21(4), pp. 243-256,

Fels, D., Richards, J., Hardman, J., & Lee, D. (2006). Sign language Web pages.*American Annals of the Deaf* , 151 (4), 423–433.

Frost, A. (2006). The Frost Village. Retrieved April 10, 2015, from http://www.frostvillage.com/.

Garcia B. &Sallandre M.‑A. (2014). *Reference resolution in French Sign Language*. In P. CabredoHofherr& A. Zribi‑Hertz (eds.) “Crosslinguistic studies on noun phrase structure and reference. Syntax and semantics series” (vol 39). Brill, Leiden: 316-364.

Gillot D. (1998). *Le droit des sourds : 115 propositions*. Report from House Reppresentative D. Gillot to French Prime Minister L. Jospin [http://www.ladocumentationfrancaise.fr/rapports-publics/984001595-le-droit-des-sourds-115-propositions-rapport-au-premier-ministre/ accessed 05/01/2015]

Goldin‑Meadow S. (2001). How do profoundly deaf children learn to read? In *Learning. Disabilities Research and Practice*, 16(4): 222-229

Hanke, T. (2004), "HamNoSys - representing sign language data in language resources and language processing contexts." In: Streiter, Oliver, Vettori, Chiara (eds): LREC 2004, Workshop proceedings : Representation and processing of sign languages. Paris : ELRA, 2004, - pp. 1-6.

Kipp, M., Nguyen, Q., Heloir, A.,&Matthes, S. (2011). Assessing the deaf user perspective on sign language avatars. In The proceedings of the 13th international ACM SIGACCESS conference on Computers and accessibility (pp. 107-114). ACM.

Lewis M.P., Simons G.F. &Fennig C.D. (eds.). 2015. Ethnologue: Languages of the World, 18th edition. Dallas, Texas: SIL International. [http://www.ethnologue.com/ accessed 05/01/2015]

Lewis, C., &Rieman, J. (1993). Task-centered user interface design: A practical introduction. University of Colorado, USA.

Norman, D., & Draper, S. (1986). *User Centered System Design*. Mahwah, NJ: Lawrence Erlbaum Associates.

Ong, S. C., &Ranganath, S. (2005). Automatic sign language analysis: A survey and the future beyond lexical meaning. *Pattern Analysis and Machine Intelligence, IEEE Transactions on*, 27(6), 873-891.

Perfetti, C., &Sandak, R. (2000). Reading optimally builds on spoken language. *Journal of Deaf Studies and Deaf Education* , 5 , 32–50.

Prensky, M. (2001). Digital natives, digital immigrants part 1. *On the horizon*, *9*(5), 1-6.

Prillwitz, S., Leven, R., Zienert, H,.Hanke, T., Henning, J. (1989). Hamburg Notation System for Sign Languages: an introductory guide, HamNoSys version 2.0.Hamburg, D: Signum Press.

Roberts, V., &Fels, D. (2006). Methods for inclusion: Employing think aloud protocols in software usability studies with individuals who are deaf. *International Journal of Human Computer Studies* , 64 (6), 489–501.

Salomoni, P., Mirri, S., Ferretti, S., &Roccetti, M. (2007). Profiling learners with special needs for custom e-learning experiences, a closed case?.In Proceedings of the 2007 international cross-disciplinary conference on Web accessibility (W4A) (pp. 84-92). ACM.

Slevinski, S.E., Jr. (2010a). International SignWriting Alphabet 2010 HTML Reference. Retrieved November 10, 2013, fromhttp://www.signbank.org/iswa.

Slevinski, S.E., Jr. (2010b). SignMaker. Retrieved April 10, 2015, fromhttp://www.signbank.org/signpuddle2.0/signmaker.php?ui=1&sgn=63.

Stokoe W.C. (1960). *Sign Language structure: an outline of the visual communication systems of the American Deaf*. Studies in linguistics: Occasional papers (No. 8). Buffalo (NY), Department of Anthropology and Linguistics, University of Buffalo..

Stokoe, W. C. (2005). Sign language structure: An outline of the visual communication systems of the American deaf. Journal of deaf studies and deaf education, 10(1), 3-37.

Sutton V. (1995). *Lessons in SignWriting: textbook, & workbook*. Deaf Action Committee for Sign Writing, La Jolla (CA, USA).

Sutton, V. (1977). Sutton Movement Shorthand: Writing Tool for Research. In W. C. Stokoe (Ed.), Proceedings of the First National Symposium on Sign Language Research & Teaching (pp. 267–296). Chicago, IL: Department of Health, Education and Welfare.

Sutton, V. (2010).SignWriting For Sign Languages - SignBank. Retrieved June 25, 2014, from http://www.signbank.org/.

Sutton, V. (2012). ASL Wikipedia Project. Retrieved June 25, 2014, from http://ase.wikipedia.wmflabs.org/wiki/Main Page.

Wilcox, S., & Wilcox, P. P. (1997).*Learning to see: Teaching American Sign Language as a second language*. Gallaudet University Press.

Wixon, D., Holtzblatt, K., & Knox, S. (1990). Contextual design: an emergent view of system design. In J. Carrasco & W. J. (Eds.), Proceedings of the SIGCHI conference on Human factors in computing systems: Empowering people (pp. 329–336). New York, NY: ACM.

Wolfe, R., Efthimiou, E., Glauert, J., Hanke, T., McDonald, J., &Schnepp, J. (2015). Special issue: recent advances in sign language translation and avatar technology. *Universal Access in the Information Society*, 1-2.